# SPACE CHARGE EFFECTS IN BUNCH SHAPE MONITORS

A.V.Feschenko, V.A.Moiseev
Institute for Nuclear Research, Moscow 117312, Russia

*Abstract*

The operation and parameters of Bunch Shape Monitors using coherent transformation of time structure of an analyzed beam into a spatial one of low energy secondary electrons emitted from a wire target is influenced by the characteristics of a beam under study. The electromagnetic field of a bunch disturbs the trajectories of secondary electrons, thus resulting in a degradation of phase resolution and in errors of phase position reading. Another effect is the perturbation of the target potential due to the current in the wire induced by a bunch as well as due to current compensating emission of the secondary electrons. The methods, the models and the results of simulations are presented.

## 1 INTRODUCTION

Bunch Shape Monitors (BSM) are used to measure longitudinal microstructure of the accelerated beam in a number of accelerators [1-4]. The principle of operation of the BSMs is based on the coherent transformation of a longitudinal distribution of charge of the analyzed beam into a spatial distribution of low energy secondary electrons through transverse RF modulation.

Typically the phase resolution of the detectors is about 1° at the frequencies of hundreds MHz. The resolution is determined by a number of parameters. The most complicated effects are due to the influence of electromagnetic fields of the analyzed beam. The fields disturb the trajectories of the electrons thus resulting in degradation of accuracy of the measurements. This effect was estimated earlier [5-7] but for extremely simplified model and detector geometry. We studied the effect for the typical geometry (fig.1) of the existing detectors and analyzed the motion of the electrons through the whole electron line from target 1 to the plane of electron collector 2.

Another effect is the perturbation of the potential of the target due to the current in the wire induced by a bunch as well as due to the current compensating emission of the secondary electrons. The model of a transmission line is used for estimating the effect.

## 2 DESCRIPTION OF THE MODEL

The motion of the electrons inside chamber 3 is analyzed for the 3D geometry. Downstream of collimator 4 a 2D model is used. Target 1, target holders 5 and the platform 6 are at the HV negative potential $U_{t\,arg}$. Chamber 3 is at zero potential.

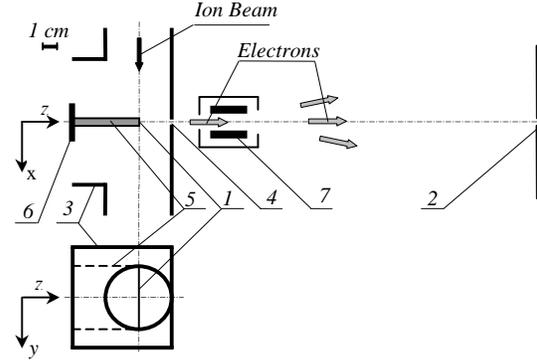

Figure 1: Geometry of the Bunch Shape Monitor.

The field inside the chamber satisfies the Poisson equation:

$$div\vec{E}(\vec{r},t) = \frac{\rho(\vec{r},t)}{\varepsilon_0}, \qquad \phi_\Gamma = f(\vec{\Gamma},t), \qquad (1)$$

where $\rho(\vec{r},t)$ is a charge density in the bunch of the analyzed beam at the moment of time $t$, $\phi_\Gamma$ is a boundary potential generally depending on time. One can split the problem (1) into three independent problems to find the fields $\vec{E}_1, \vec{E}_2$ and $\vec{E}_3$ ($\vec{E} = \vec{E}_1 + \vec{E}_2 + \vec{E}_3$).

Problem 1: $\qquad div\vec{E}_1(\vec{r}) = 0, \qquad \phi_1 = f_0(\vec{\Gamma}) \qquad (2)$

The field $\vec{E}_1$ can be found from a solution of the Laplace equation for the potential $\phi_1(\vec{r})$ without a beam:
$\vec{E}_1(\vec{r}) = -grad\phi_1(\vec{r})$

Problem 2: $\quad div\vec{E}_2(\vec{r},t) = 0, \quad \phi_2(\vec{\Gamma},t) = f_2(\vec{\Gamma},t) \qquad (3)$

We assume this process to be a quasi stationary one and $|\vec{E}_2| << |\vec{E}_1|$. In this case the potential satisfies the Laplace equation and $\vec{E}_2(\vec{r},t) \approx -grad\phi_2(\vec{r},t)$. Formulation of this problem is an attempt to take into account the effects of distortion of the boundary potential. The distortions of the target voltage are estimated below but are not taken into account in the simulation process.

Problem 3: Generally the bunch generates both electric and magnetic fields and a complete system of Maxwell equations must be solved. To simplify the problem we consider it to be electrostatic in the reference frame moving with the bunch. The magnetic field due to the charges located on the moving boundaries in this frame is neglected. With this assumptions for each fixed moment

of time $t_0$ the electric field in the beam frame $\vec{E}_{03}$ can be found from the Poisson equation

$$div\vec{E}_{03}(\vec{r}_0,t_0) = \frac{\rho(\vec{r}_0,t_0)}{\varepsilon_0}, \qquad f_{03}(\vec{\Gamma}_0,t_0) = 0 \qquad (4)$$

The subscript "0" indicates that the beam frame is considered. The electric and magnetic fields in the laboratory frame can be found by Lorenz transformations.

The equations (2) and (4) were solved numerically for the mesh 0.5mm×0.5mm×0.5mm (in the laboratory frame). Near the target in the region where the motion of the electrons is of interest the radial component of the electric field is approximated by the function $E_r(x,y,z) \approx \frac{K_1(y)}{r}$ for the problem 1 and $E_r(x,y,z) \approx K_3(y)e^{-K_4(y)r^2}$ for the problem 3 [8]. Here $r$ is the distance from the target center. The functions $K_1(y)$ and $K_4(y)$ were selected to satisfy the condition $W = -eU_{targ}$ for the fixed position of the bunch, where $W$ is the energy of electrons passing through slit 4 and emitted from the target with the zero energy.

Phase resolution of the detector is defined as $\Delta\varphi = \frac{\Delta X_L}{X_{max}}$. To avoid mixture of the effects $\Delta X_L$ in our analysis is considered to be the size of the focused electron beam in the plane of electron collector 2 with the RF deflecting field in deflector 7 off. $X_{max}$ is the amplitude of the displacement of the electrons. For simplicity we assume $\Delta X_L \approx 2\sigma$, where $\sigma$ is rms size of the focused electron beam. Due to space charge the size $\Delta X_L$ increases thus resulting in a degradation of $\Delta\varphi$. Another effect is the changing of the average position of the focused electron beam $\delta X_L$. This effect is the reason for the error of phase reading $\delta\varphi = \frac{\delta X_L}{X_{max}}$.

Modulation of the electron velocity in $y$ direction also can result in distortion of the results of the measurement because of possible displacement of the beam outside the active area of the electron detector.

To estimate the distortion of the potential of the target it was considered as a transmission line (fig. 2).

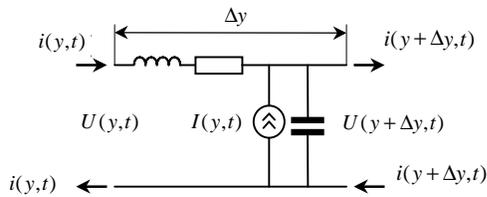

Figure 2: Equivalent circuit of the target.

Charge density on the conducting surface can be written as $\sigma = E_n \cdot \varepsilon_0$, where $E_n$ is the component of electric field perpendicular to the surface. Linear density of the charge in the wire: $q(y,t) = \int_0^{2\pi}\sigma r_t d\alpha$, where $r_t$ is a target radius. The relation of the changing of the charge distribution due to motion of the bunch and the current in the wire can be written as: $\frac{\partial q(y,t)}{\partial t} = \frac{\partial i(y,t)}{\partial y}$. For the transmission line model the voltage $U(y,t)$ due to the current $i(y,t)$ satisfies the equation

$$\frac{\partial U(y,t)}{\partial y} = L\frac{\partial i(y,t)}{\partial t} + Ri(y,t) \qquad (5)$$

Taking into account that the typical transverse dimensions of the target are much smaller than those of the holders 5 the boundary conditions for $y = \pm l_{targ}/2$, where $l_{targ}$ - is the length of the target wire, can be set to zero. Zero moment of time can be selected to correspond to the position of the bunch at the entrance of the chamber and the initial condition also can be set to zero.

The emission current is represented by the current source $I(y,t)$. One can show that the voltage $U$ in the transmission line due to this current satisfies the equation

$$\frac{\partial^2 U}{\partial y^2} - LC\frac{\partial^2 U}{\partial t^2} - RC\frac{\partial U}{\partial t} = RI + L\frac{\partial I}{\partial t} \qquad (6)$$

The boundary and initial conditions can be the same as previously with the zero moment of time corresponding to the middle point between bunches.

The model of a transmission line cannot be considered as a rigorous one. We did not use it for problem 2. However we believe it to be useful for estimationg the perturbations of the target voltage and for making a conclusion about the applicability of problems 1 and 3.

## 3 RESULTS OF SIMULATIONS

The simulations have been done for the 3D Bunch Shape monitor [3] installed in the CERN Linac-2 ( $I$ =150 mA, $\sigma_\varphi$=15°, $\sigma_y$=3.5 mm, $\sigma_z$=2 mm, $f$=200 MHz, $U_{targ}$=-10 kV, $X_{max}$=27 mm).

The trajectories of the electrons, emitted at different phases, in the space between the target and collimator 4 are presented in fig. 3.

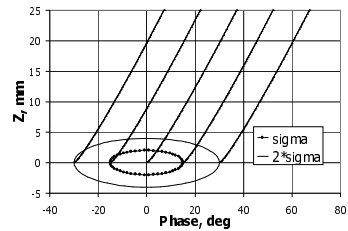

Figure 3: Trajectories of electrons in the beam frame emitted at different moments of time.

The difference of electron energy with and without space charge along the trajectory for the electrons corresponding to the head and the center of the bunch is given in fig. 4. Modulation of energy of the electrons at the collimator 4 and time of flight from the target to RF deflector 7 is shown in fig. 5.

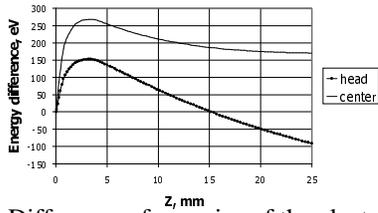

Figure 4: Difference of energies of the electrons with and without space charge.

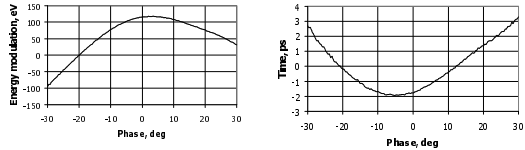

Figure 5: Modulation of electron energy and time of flight.

The behavior of phase resolution and error of phase reading is shown in fig. 6.

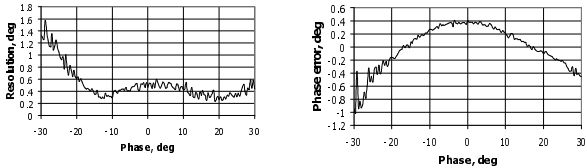

Figure 6: Phase resolution and error of phase reading.

The displacement of the electrons in the plane of electron collector 2 in the $y$ direction, due to space charge with respect to their position without space charge, is presented in fig. 7.

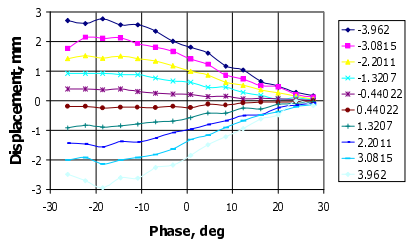

Figure 7: Displacement of the electrons due to space charge for different initial coordinates $y$.

Perturbation of the target voltage due to the current induced by the bunch and calculated for $L = 1.24 \cdot 10^{-6}$ H/m, $C = 8.93 \cdot 10^{-12}$ F/m and $R = 61$ Ohm/m (the parameters of the equivalent coaxial transmission line with the diameter of the inner and outer conductors of 0.1 mm and 50 mm correspondingly at 1 GHz) is shown in fig. 8. The solution of equation (6) also gives a bipolar shape of perturbation. The magnitude of the perturbation for the experimentally estimated emission current of 250 µA does not exceed ± 5 µV.

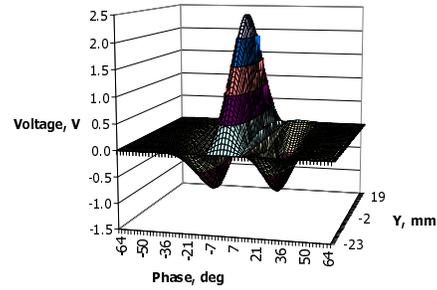

Figure 8: Perturbation of the target voltage due to the current induced by the bunch.

## 4 CONCLUSION

The limits of the report do not enable a thorough study of the effects to be presented. Although each particular set of parameters requires special calculations, nevertheless some general rules can be mentioned. Thus increasing the target potential evidently decreases all the space charge effects. Although decreasing the transverse dimensions of the beam increases the space charge forces, nevertheless it leads to the shrinking of the region of the intensive interaction of the electrons with the bunch and as a result decreases the influence of space charge. Decreasing the longitudinal bunch size initially gives rise to the space charge effects. However further decreasing results in decreasing the time of interaction of the electrons with the bunch and hence in decreasing the final effects. The effects of voltage perturbation are essential for short bunches and for some practical parameters are dominating.

## REFERENCES

[1] P.N.Ostroumov et al. Proc of the XIX Int. Linear Acc. Conf., Chicago, 1998 pp.905-907.A
[2] A.V.Feschenko et al. Proc. of the 1997 PAC, Vancouver, 1997, pp.2078-2080.
[3] S.K.Esin et al. Proc. of the XVIII Int. Linear Acc. Conf., Geneva, 1996, pp.193-195.
[4] S.K.Esin et al. Proc. of the 1995 PAC and International Conf. on High-Energy Accelerators, Dallas, 1995, pp.2408-2410.
[5] V.Vorontsov, A. Tron. Proc of the 10[th] All-Union Workshop on Part. Acc. Dubna 1986, V.1, pp.452-455 (in Russian).
[6] A.Tron, I.Merinov. Proc. of the 1997 PAC, Vancouver, 1997, V.2, pp.2247-2049.
[7] E.McCrory et al Proc. of the XVIII Int. Linear Acc. Conf., Geneva, 1996, pp.332-334.
[8] C.Badsel, A.Lengdon. Phisika Plazmy I chislennoe Modelirovanie, Energoatomizdat, Moscow, 1989 (in Russian).